\documentclass[twocolumn,english,showpacs]{revtex4}
\usepackage{graphicx}
\usepackage{epsfig}
\usepackage{amsmath}
\usepackage{amssymb}

\usepackage{hyperref}
\usepackage{units}

\begin{document}

\title{Tomography by noise}
\author{G. Harder$^1$, D. Mogilevtsev$^2$, N. Korolkova$^3$ and Ch. Silberhorn$^{1,4}$}

\affiliation{ $^1$Applied Physics, University of Paderborn,
Warburgerstrasse 100, 33098 Paderborn, Germany;
\\
$^2$Institute of Physics, Belarus National Academy of Sciences,
F.Skarina Ave. 68, Minsk 220072 Belarus;
\\
$^3$School of Physics and Astronomy, University of St Andrews,
North Haugh, St Andrews KY16 9SS, UK;
\\
$^4$Max Planck Institute for the Science of Light,
Guenther-Scharowsky-Strasse 1/ Building 24, 91058 Erlangen,
Germany }

\begin{abstract}
We present an efficient and robust method for the reconstruction
of photon number distributions by using solely thermal noise as a
probe. The method uses a minimal number of pre-calibrated quantum
devices, only one on/off single-photon detector is sufficient.
Feasibility of the method is demonstrated by the experimental
inference of single-photon, thermal and two-photon states. The
method is stable to experimental imperfections and provides a
direct, user-friendly quantum diagnostics tool.
\end{abstract}

 \maketitle

\textit{Introduction. -} Ultimately, the quantum tomography is the
most comprehensive tool available for a researcher. Indeed, by
inferring the quantum state we have a possibility to predict
results of any possible measurement. From its birth in 1989
\cite{vogel}, quantum tomography has made an enormous progress
\cite{all,special}. Now even such fragile quantum objects as
``Schr\"{o}dinger cats" made of photons are diagnosed and
reconstructed \cite{cats}. However, the most precise tool requires
the most precise tuning. Generally, the quantum reconstruction
schemes require precise calibration of the measurement set-up
together with minimization of noise and losses. For example, one
of the most established tomographic tools for electromagnetic
field states, the quantum homodyne tomography requires more than
$50\%$ overall detection efficiency \cite{dariano95}. Also, rather
low respective phase noise of the probe and signal fields is
essential for the scheme to work.

In this Letter we present a quantum tomography scheme that
actually relies on the noise to collect data sufficient for the
state reconstruction. Furthermore, the data is collected by using
merely one on/off detector, where the ability to distinguish the
number of the input photons is not required. The essence of the
scheme is simple: the signal mixed with the thermal noise impinges
on the detector. Varying intensity of the noise, we can build up
the set of measurements sufficient for the inference of diagonal
elements of the signal density matrix. The reconstruction can be
done even for quite low detection efficiencies on the level of
10$\%$ . An important feature of our scheme is the minimization of
resources. Even the simplest of the conventional schemes using one
on/off avalanche photo-detector still requires a number of
pre-calibrated absorbers or beam-splitters \cite{mog98}. With
increasing signal intensity, this number increases dramatically.
The schemes based on time-multiplexing or space-multiplexing
similarly involve a considerable amount of pre-calibration
\cite{loop,timemultiplex}. Additionally, they assume that the
signal does not contain photon number contributions beyond the
number of multiplexing channels. In contrast, in our scheme the
detector itself can be used to determine the temperature of noise,
thus avoiding the necessity to have any other pre-calibrated devices.
Moreover, there is no restriction to the low dimensional Hilbert
space corresponding to low input photon number pre-defined by the
detector. Our scheme can be generalized to enable a complete reconstruction
of the signal state density matrix by mixing the signal with the coherent field.

\begin{figure*}[ht]
\begin{center}
\includegraphics[width=1.0\linewidth]{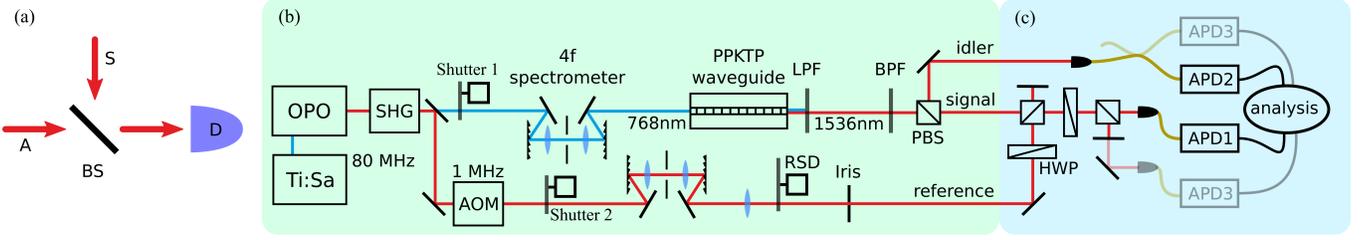}
\caption{(a) The sketch of the measurement scheme: the signal A is mixed with the probe S
on the beam-splitter (BS) and impinges on the bucket detector D.
(b)  State preparation setup. Pulsed light at telecom
wavelengths is generated in a Ti:Sapph pumped optical parametric
oscillator (OPO) and frequency doubled by second harmonic
generation (SHG). Part of the light is separated for later use as
a reference field. The repetition rate of the reference is lowered
by an acousto-optical modulator (AOM). The 4-f spectrometer
tailors the spectral width of the pump beam to achieve spectral
decorrelation. The PDC state is generated inside the periodically
poled KTP waveguide. The pump is separated by a long pass filter
(LPF). A bandpass filter (BPF) is used to suppress background
outside the PDC spectrum. Finally signal and idler are separated
at a polarizing beam splitter (PBS). The reference field is also
spectrally tailored by a 4-f setup. Pseudo-thermal light is
generated by a rotating speckle disk (RSD) followed by irises. (c)
Measurement setup. Signal and reference are overlapped at a
PBS-HWP-PBS combination, which effectively constitutes a variable
BS adjusted to a splitting ration of $90/10$. The power
of the reference is controlled by the first HWP.  Probed signal
and idler are coupled into single mode fibers and impinge onto two
avalanche photo diodes (APDs) (Id Quantique id201 at a repetition
rate of 1{MHz} with a gate width of about 2.5{ns}).
A third APD is used either in the idler beam to herald two photon
states or in the second output port of the variable BS
to estimate the mode overlap between signal and reference by Hong
Ou Mandel interference.} \label{fig:setup}
\end{center}
\end{figure*}

\textit{The scheme. -}  We first demonstrate the feasibility of our
scheme in its simplest configuration. The goal is to infer
diagonal elements of the signal density matrix, $\rho_{mm}$, in
the Fock-state basis, $|m\rangle$. The probability of registering
a signal is generally given as
\begin{equation}
p_j=\sum\limits_{m=0}^{N}\Pi_{jm}\rho_{mm}, \label{born}
\end{equation}
where  the elements $\Pi_{jm}=\langle m|\Pi_{j}|m\rangle$ are
related to the $j$th element of positive valued operator measure
(POVM), $\Pi_{j}$, which describes a measurement performed on the
signal. $N+1$ is the dimension of the subspace of all possible
signal states. We have only one on/off detector and we use
different thermal probe states to generate different POVM
elements. Let us suppose that
the probe completely overlaps with the signal at the detector,
which has a detector efficiency $\eta$. If we now register ``no
click" events, we obtain POVM operator matrix elements for such a
measurement \cite{perina,rockower}:
\begin{equation}
\Pi_{jm}=y_j(1-y_j{\eta})^m, \label{p_simple}
\end{equation}
where $y_j={1}/(1+\eta{n}_j)$ and $n_j$ is the mean photon number
of the probe thermal state (TS). The matrix with elements $(x_j)^m$ is
always non-degenerate for $N+1$ different values of $x_j$ and
$m=0,1\ldots N$ (it is the Vandermonde matrix
\cite{horn}). Since we can represent the system (\ref{born}) as
$p_j/y_j=\sum\limits_{m=0}^{N}(1-y_j{\eta})^m\rho_{mm}$, it means
that using TS probes provides us with measurements, which
should provide enough information to reconstruct elements $\rho_{mm}$.
To collect the necessary data,  only one pre-calibrated
on/off detector is needed and it is sufficient to change the
probe arbitrarily. When the signal is blocked,  the
average number of photons, $n_j$, in the probe can be measured.

In practice, instead of the simplest scheme (\ref{p_simple}), we
have mixed the probe with the signal using a
beam-splitter (BS) (see the scheme (a) in Fig.(\ref{fig:setup})). For two imperfectly overlapping fields, the signal
$a$ and the probe $b$, interfering on BS and
afterwards impinging on the detector, the probability to register
``no click" is given by \cite{laiho}:
\begin{eqnarray}
\nonumber p_j= {\rm Tr} \{
:\exp\{-\eta(Ta^{\dagger}a+(1-T)b^{\dagger}b+
\\
x(a^{\dagger}b+b^{\dagger}a))\}:\rho\sigma_j \},\label{general}
\end{eqnarray}
where $a^{\dagger}$, $a$ and $b^{\dagger}$, $b$ are the creation
and annihilation operators of signal and probe modes;  $\sigma_j$
is the density matrix of the $j$th probe field; $T$ is the
transmissivity of BS; $x=\sqrt{\mu T(1-T)}$, $\mu$
is the overlap parameter; $::$ denotes the normal ordering
operator. For the perfect overlap, Eq.(\ref{general}) results in a
straightforward relation:
\begin{equation}
\Pi_{jm}=\sum\limits_{n,k,l=0}^{N}(1-\eta)^k\sigma_{nj}|U^{kl}_{mn}|^2.
\label{uoverlap}
\end{equation}
Quantities $\sigma_{nj}=(n_j)^n/(1+n_j)^{n+1}$ are diagonal matrix
elements of the $j$th probe TS. The operator $U$
describes the rotation performed by BS. It has the
following matrix elements in the Fock-state basis:
\begin{eqnarray}
\nonumber
U^{kl}_{mn}=\sqrt{k!l!m!n!}\sum\limits_{g=0}^k\sum\limits_{h=0}^l\frac{t^{g+h}r^{k+l-g-h}(-1)^{k-g}}{g!h!(k-g)!(h-l)!}\times
\\
\delta_{m,l+g-h}\delta_{n,k+h-g}.
\end{eqnarray}
Here $t=\sqrt{T}$ and $r=\sqrt{1-T}$. For the zero-temperature noise, $n_j=0$, Eq.(\ref{uoverlap}) gives
\begin{equation}
p_{\rm{signal}}=\sum\limits_{k=0}^{N}(1-T\eta)^k\rho_{kk}.
\label{psignal}
\end{equation}
Now let us represent the probe TS as a mixture of
coherent states, $|\alpha\rangle$, \cite{perina}:
$\sigma_{j}=\frac{1}{\pi n_j}\int d^2\alpha
\exp\{-|\alpha|^2/n_j\}|\alpha\rangle\langle\alpha|$.
In the case of the perfect overlap, $p_j$ for probe TS can be expressed through the
probability of ``no clicks" for the coherent probe (given in
Ref.\cite{laiho}):
\begin{eqnarray}
\nonumber p_j= \frac{1}{\pi {n}_j}\int d^2\alpha
\exp\{-|\alpha|^2/n_j\}\times \\
 \langle :\exp\{-\eta
T(a^{\dagger}+\nu\alpha^*)(a+\nu\alpha)
\}:\rangle_a,\label{nooverlap1}
\end{eqnarray}
where $\nu=\sqrt{(1-T)/T}$. Notice, that this formula is
equivalent to the expression for $p_j$ given by POVM elements
(\ref{uoverlap}) for $N\rightarrow\infty$. The POVM elements for
an imperfect overlap can be derived representing
Eq.(\ref{general}) for $p_j$ for the
 probe TS in the form similar to Eq.(\ref{nooverlap1}):
\begin{eqnarray}
\nonumber p_j= \frac{{\bar n}_j}{\mu n_j}\frac{1}{\pi {\bar
n}_j}\int d^2\alpha \exp\{-|\alpha|^2/{\bar n}_j\}\times
\\
\langle :\exp\{-\eta T(a^{\dagger}+\nu\alpha^*)(a+\nu\alpha)
\}:\rangle_a.\label{overlap1}
\end{eqnarray}
The quantity
${\bar n}_j={\mu n_j}/(1+(1-\mu)(1-T)\eta n_j)$.
Comparing the expression (\ref{overlap1}) with Eqs.
(\ref{nooverlap1}), (\ref{uoverlap}) for the perfect overlap, we
obtain the relation for the POVM elements in case of the imperfect
overlap:
\begin{equation}
\Pi_{jm}^{\rm overlap}=\frac{{\bar n}_j}{\mu
n_j}\sum\limits_{n,k,l=0}^{N}(1-\eta)^k{\bar
\sigma}_{nj}|U^{kl}_{mn}|^2, \label{overlap2}
\end{equation}
where ${\bar \sigma}_{nj}$ are the diagonal matrix elements of TS with the average number of photons ${\bar n}_j$. Eq.
(\ref{overlap2}) points to a number
of important conclusions. First of all, for the zero overlap, the
``modified" average number of photons is also zero, ${\bar
n}_j=0$. As follows from Eqs.(\ref{psignal}), (\ref{overlap2}),
the resulting ``no click" probability factorizes,
$p_j(\mu\rightarrow0)\rightarrow{p_{\rm{signal}}}p_{\rm{term}}$,
where $p_{\rm{term}}=1/(1+(1-T)\eta n_j)$ is the ``no click"
probability for the probe TS with the vacuum instead of the
signal.  For a weak probe, when $(1-\mu)(1-T)\eta n_j << 1$, the
actual situation can be modelled by having two probe modes, the
one completely overlapping with the signal with average number of
photons equal to $\mu n_j$, and the non-overlapping one with
average number of photons equal to $(1-\mu) n_j$. When the probe
is strong, $(1-\mu)(1-T)\eta n_j
>> 1$,  part of the probe actually interfering with the signal remains
constant, ${\bar n}_j\approx \mu /(1-\mu)(1-T)\eta$. In other
words, too strong probe will wash out effects  of
interference and destroy a possibility to reconstruct the signal.
The optimal regime is the moderate levels of the probe TS.

It should be noticed that our set-up can be easily generalized for the complete state reconstruction.
Coherently shifting the signal with the amplitude $\alpha$, one can reconstruct
the set of following quantities: $\langle m|D(\alpha)\rho D^{\dagger}(\alpha)\ |m \rangle$, where the coherent shift operator is $D(\alpha)=\exp\{\alpha a^{\dagger}-\alpha^* a\}$. For an $N$-dimensional density matrix of the signal, it is sufficient to have
$N$ different settings of the coherent shift and $N$ TS to infer the complete density matrix (for the procedure see, for example, Refs.\cite{ourprl2006}). To realize this generalization with our set-up, apart from the one additional fixed BS, one needs only to have a pre-calibrated phase-shifter to change a relative phase of the added coherent field.

\textit{Setup. -} For the signal state generation we employ a
type-II parametric downconversion (PDC) source in a periodically
poled KTP waveguide. The source is characterized in detail in
\cite{harder_optimized_2013}. It produces spectrally nearly
decorrelated PDC states such that heralded states have a high
purity above $80\%$. Furthermore, being a waveguide source, it
allows for efficient coupling into single mode fibers. A scheme of
the full setup is shown in Fig. \ref{fig:setup}. To simulate a noise source with TS photon number statistics,
we generate pseudo-thermal light using a rotating speckle disk.
For each position of the speckle disk, a random interference
pattern is created. After spacial filtering by irises and the
final fiber incoupling, the intensity shows an exponential, hence
thermal, probability distribution. We verify the thermal
statistics by measuring the mean photon number for different
positions of the disk as well as by the second-order correlation
function $g^{(2)}$. We obtain $g^{(2)}>1.9$, whereas a value of
$g^{(2)}=2$ corresponds to perfect thermal statistics and
$g^{(2)}=1$ to Poissonian statistics. The remaining coherent part
is thus very small and can be neglected. The calibration parameters of our scheme are the mode overlap
between signal and probe $\mu$ and the overall efficiency $\eta$.
To determine the mode overlap, we adjust our variable beam
splitter to 50/50 and measure a Hong-Ou-Mandel dip. The overlap is
calculated from the visibility of the dip as described in
\cite{Kaisa_HOM} to be $\mu=0.45$. The decrease from unity comes
possibly from a spectral mismatch in the 4-f setup or a spacial
mismatch while coupling into the fiber. The detection efficiency
is measured using the Klyshko scheme \cite{klyshko} from which we
obtain $\eta=0.15$. To generate a set of probe states, we rotate a
HWP (see Fig. \ref{fig:setup}) and measure the mean photon numbers
$n_j$ from counts in APD1 with a physically blocked PDC beam.

\begin{figure}[ht]
\begin{center}
\includegraphics[width=1.1\linewidth]{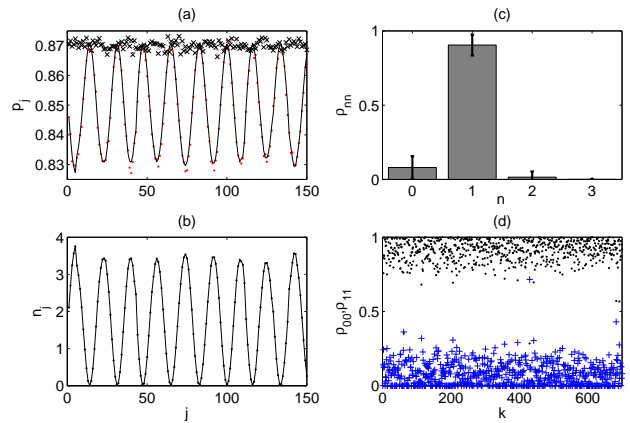}
\caption{Reconstruction of {\it the heralded single-photon state} for
the scheme parameters $\eta=0.15$ and $\mu=0.45$. $10^7$ PDC
pulses were used for each value of the reference field intensity
(a). Dots show experimentally collected data for the number of
``no clicks" with respect to the total number of pulses for the
signal overlapped with the reference beam. The oscillating
behavior comes from the fact that a HWP is used to change the
reference beam power at each measurement point. Crosses depict
the same probability of ``no clicks" on the APD for the signal alone;
solid line shows probabilities estimated by Eq.(\ref{overlap2})
for the result shown in Fig.~\ref{fig1}(c). (b) Average number of
thermal photons, $n_j$, of the probe for the
data of (a). (c) Experimentally inferred $\rho_{nn}$ of the
heralded single-photon state. (d) Experimentally estimated values
of vacuum (crosses) and single-photon (dots) components of the
signal obtained via bootstrapping the data shown in the panel (a).
} \label{fig1}
\end{center}
\end{figure}

\begin{figure}[ht]
\begin{center}
\includegraphics[width=1.1\linewidth]{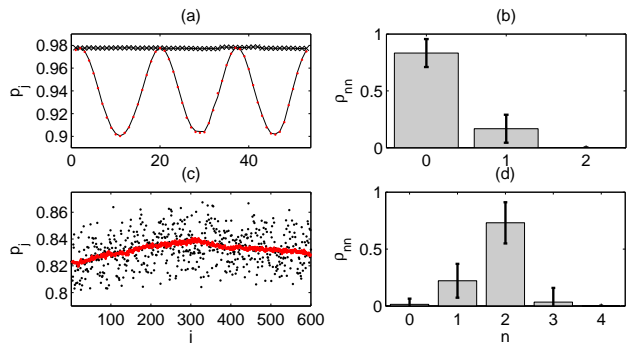}
\caption{ (a, b) Reconstruction of {\it the thermal state}. (a) Dots
show experimentally collected data for the number of ``no clicks"
with respect to the total number of pulses for the unheralded signal
overlapped with the probe. Crosses depict the relative
number of ``no clicks" on the APD for the signal alone; solid line
shows probabilities estimated by Eq.(\ref{overlap2}) for the
average reconstruction result shown in Fig.~\ref{fig2}~(b). (c, d)
Reconstruction of  {\it the  two-photon state}. (c) Dots show experimentally
collected data for the number of ``no clicks" with respect to
the total number of pulses for the heralded two-photon signal
without the reference; solid line shows experimentally collected
data for the number of ``no clicks" in respect to the total
number of pulses for the heralded single-photon signal without the
reference. (d) Experimentally inferred  $\rho_{nn}$ of the
heralded two-photon state obtained accounting
for the efficiency drift. The scheme
parameters are as in Fig.~\ref{fig1}; 600 different settings of
the reference field were taken. } \label{fig2}
\end{center}
\end{figure}

\textit{Results. -} Fig.~\ref{fig1} shows the results of
reconstruction for the heralded \textit{single-photon state}
generated by the scheme depicted in Fig.~\ref{fig:setup}.  Total
150 measurement points were used for the inference. The
reconstruction was done using least-square estimation with
non-negativity constraints \cite{our njp 2013}. The detection
efficiency $\eta=0.15$ and the overlap $\mu=0.45$ were assumed.
Fig.~\ref{fig1}~(d) visualizes the estimated values of vacuum and
single-photon components of the signal obtained via bootstrapping
the data \cite{bootstrap}.
Our reconstruction procedure for the single-photon state gives the following
value of the single-photon component
$\rho_{11}\approx 0.905\pm0.07$. This estimate conforms well with the result of the recent work
\cite{harder_optimized_2013} where the same source was used,
 demonstrating high quality of the reconstruction. Also, quite similar result was obtained with
 the same source using the "data pattern" reconstruction method \cite{arxivzdenek}.
Fig.~\ref{fig2} shows experimentally obtained data for the
heralded \textit{two photon state} and the \textit{thermal
state} \cite{pdc thermal}.  In
Fig.~\ref{fig2}(a) only a part of the measured data is shown. Here,
when varying the probe intensity, 600 different values of
reference field intensity were taken. The average photon-number
distribution shown in Fig.~\ref{fig2}~(b) is close to the thermal
with the average number of photons equal to $0.17$. Relatively
large values of variances might be explained by the fact that the
signal field intensity was significantly higher in difference to
the case shown in Fig.~\ref{fig1}. As a consequence, parameters of
the measurement set-up were not as stable. In particular, the
detection efficiency was drifting, so that an efficiency drift of
about $15\%$ was registered. This deviation might result from a
drift in the fiber incoupling efficiency due to instabilities of
the setup over the measurement time. The data for the heralded
two-photon state is affected in a similar way as can be clearly
seen also in Fig.~\ref{fig2}~(c). Here we show experimental data
for both heralded two-photon signal (dots) and heralded
single-photon signal (solid line) not mixed with the reference.
One of the powerful features of our method is the possibility to
account for these deviations, by assuming varying efficiency. For
the estimation of the detection efficiency, we need to use the
data for the signal state not mixed with the reference. For
example, if we take the single-photon state, use
Eq.~(\ref{psignal})  and the experimentally measured probability
$p_{\rm signal}$ as shown in Fig.~\ref{fig1}, we can compute the
actual values of $\eta$. The drift in the detection efficiency
$\eta$ is reflected in the varying value of $p_j$ for the heralded
single-photon signal without the reference, as depicted by the
solid (red) line in Fig.~\ref{fig2}~(c). Ideally, this should be a
straight line, as it is approximately  for the data set with the
low field intensity, used for the single-photon state
reconstruction ($p_j$ in Fig.~\ref{fig1}~(a), solid line). For the
data set with the higher field intensity, as used for the
two-photon reconstruction, this is not the case anymore
(Fig.~\ref{fig2}~(c)). To account for this, we incorporated the
calculated actual efficiency values in the expression for the POVM
elements  (\ref{overlap2}) when inferring $\rho_{nn}$  for  the
generated two-photon state (Fig.~\ref{fig2}~(d)). The obtained results for the
two-photon signal are quite similar to obtained recently with the same source using
the "data pattern" method \cite{arxivzdenek}. It should be emphasized that deviations of
obtained data do not lead to reconstruction artifacts in our
scheme. For example, the vacuum component of the reconstructed
signal remains very low despite a rather noisy character of the
data. Also, the result of reconstruction does unambiguously show
that despite low efficiencies of the detection, the scheme
produces states with large two-photon component. All these feature are preserved even if no correction for varying detection
efficiency is performed for the two-photon state (Fig.~\ref{fig2}~(d)), although the relative errors are much higher then.

\textit{Conclusions. -}  We have demonstrated both theoretically
and experimentally that reconstruction by noise is indeed feasible
and provides a lucid, robust tomographic tool. By merely mixing
the signal with the thermal noise and measuring statistics of the
resulting field on the on/off detector, we can collect data
sufficient for inferring photon-number distributions of different
signal fields. Our reconstruction scheme required only a minimum
number of pre-calibrated devices operating on the single-photon
level. For collecting data, only a single on/off detector and a
fixed ratio BS were used. The reference field (thermal
light) was calibrated using the same detector. Reconstruction of
single-photon, thermal and two-photon states was performed. The
scheme has proven to be quite robust with respect to the
noise/deviation affecting the measurement set-up. Our scheme can
be generalized to a complete tomography by adding coherent shifts
to the signal. We believe that such a scheme can become a simple,
inexpensive and efficient working tool of quantum diagnostics.
Potentially, even spectrally filtered light from such incoherent
sources as an incandescent lamp can be used for the probe.

\textit{Acknowledgements. - } N. K. acknowledges the support
provided by the A. von Humboldt Foundation and the Scottish
Universities Physics Alliance (SUPA).  The research leading to
these results has received funding from the European Community's
Seventh Framework Programme (FP7/2007-2013) under grant agreement
n$^\circ$ 270843 (iQIT). We are grateful for the support of the
International Max Planck Partnership (IMPP) for Measurement and
Observation at the Quantum Limit. This work was also supported by
NASB through the program "Convergence" (D.M.) We are very thankful to
J. Pe\v{r}ina and V. S. Shchesnovich for fruitful discussions.

\end{document}